%% file: arxiv.tex
\documentclass[compsoc,onecolumn]{IEEEtran}

\usepackage{cite}
\usepackage[pdftex]{graphicx}
\usepackage{amssymb}
\usepackage[bottom,flushmargin,hang,multiple,norule]{footmisc}
\usepackage[cmex10]{amsmath}
\usepackage{url}
\usepackage{algorithm}
\usepackage[caption=true,font=normalsize,labelfont=sf,textfont=sf]{subfig}
\usepackage{algorithmic}
\usepackage{setspace}

\newcommand{\ob}{out-of-the-box }

\newcommand{\0}{\mathbf{0}}
\newcommand{\I}{\mathbf{I}}
\newcommand{\Ncal}{\mathcal{N}}
\newcommand{\mmu}{\boldsymbol{\mu}}
\newcommand{\aalpha}{\boldsymbol{\alpha}}
\newcommand{\SSigma}{\boldsymbol{\Sigma}}
\newcommand{\W}{\mathbf{W}}
\newcommand{\A}{\mathbf{A}}
\newcommand{\X}{\mathbf{X}}
\newcommand{\x}{\mathbf{x}}
\newcommand{\y}{\mathbf{y}}

\newcommand{\w}{\mathbf{w}}

\makeatletter
\def\blfootnote{\xdef\@thefnmark{}\@footnotetext}
\makeatother

\usepackage[normalem]{ulem}
\usepackage{color}

\begin{document}
\title{Transfer Learning in Brain-Computer Interfaces}
\bibliographystyle{./sty/IEEEtran}

\author{\IEEEauthorblockA{\bf{Vinay Jayaram}\\ {\normalfont IMPRS for Cognitive and Systems Neuroscience, University of T\"{u}bingen, T\"{u}bingen, Germany \\
Department of Empirical Inference, Max Planck Institute for Intelligent Systems, T\"{u}bingen, Germany}}\\
\IEEEauthorblockA{\bf{Morteza Alamgir}\\\normalfont Department of Computer Science, University of Hamburg, Hamburg, Germany}\\
\IEEEauthorblockA{\bf{Yasemin Altun\\Bernhard Sch\"{o}lkopf\\Moritz Grosse-Wentrup}\\\normalfont Department of Empirical Inference, Max Planck Institute for Intelligent Systems, T\"{u}bingen, Germany} \\
}

\maketitle

\begin{abstract}
\input{abstract.tex}
\end{abstract}

\IEEEpeerreviewmaketitle

\maketitle

\blfootnote{Corresponding author: Vinay Jayaram (email: vjayaram@tuebingen.mpg.de)}
\input{Introduction_2.tex}

\input{Training.tex}
\input{Online.tex}

\section{Experiments}
\label{sec:exper}
We conducted two experiments with real-world data sets. The first used both the initial
multitask learning algorithm as well as the version with decomposition of spectral
and spatial features while the second only used the version with feature decomposition (hereafter referred to as FD). 
The first is an example of subject-to-subject transfer with a motor imagery dataset recorded for ten healthy subjects,
and the second is an example of session-to-session transfer for a neurofeedback paradigm recorded in a single subject with ALS. 

\input{MotorImagery}
\input{DeltaFeedback}
\input{Discussion}

\section{Conclusion}
Previous approaches to transfer learning in BCI have ignored the possibilities of knowledge transfer within the feature space, constraining themselves mostly
to spatial filtering and domain adaptation. Here, we present a method for learning that transfers knowledge from previous subjects to new ones in 
any desired spatiotemporal feature space, able to work both on its own and on top of other paradigms. Testing on both motor imagery and a novel cognitive paradigm, we find that our proposed methods better deal with both session-to-session and subject-to-subject variability as compared to simple pooling, achieving accuracies comparable to or better than single-session training with far fewer training trials. Further, this work presents a framework on top of 
which other objective functions can be used to determine priors for decision boundaries that minimize other sorts of error. Any parties interested in
trying these algorithms for themselves will find implementations of all three algorithms in MATLAB at the following website: \url{http://brain-computer-interfaces.net/}. 

\section*{Acknowledgment}

The authors would like to thank Tatiana Fomina, Christian F\"{o}rster, Natalie Widmann, Marius Klug, and Nadine Simon for their help in gathering the data
used in the second experiment.

\bibliography{bibliography.bib}

\end{document}

%% file: abstract.tex
The performance of brain-computer interfaces (BCIs) improves with the amount of available training data; the statistical distribution of this data, however, varies across subjects as well as across sessions within individual subjects, limiting the transferability of training data or trained models between them. In this article, we review current transfer learning techniques in BCIs that exploit shared structure between training data of multiple subjects and/or sessions to increase performance. We then present a framework for transfer learning in the context of BCIs that can be applied to any arbitrary feature space, as well
as a novel regression estimation method that is specifically designed for the structure of a system based on the electroencephalogram (EEG). We demonstrate the utility of our framework  and method on subject-to-subject transfer in a motor-imagery paradigm as well as on session-to-session transfer in one patient diagnosed with amyotrophic lateral sclerosis (ALS), showing that it is able to outperform other comparable methods on an identical dataset.

%% file: Introduction_2.tex
\section{Introduction}
It is often a problem in various fields that one runs into a series of tasks that appear - to a human - to be highly related to each other, yet applying the optimal machine learning solution of one problem to another results in poor performance. Specifically in the field of brain-computer interfaces (BCIs), it has long been known that a subject with good classification of some brain signal today could come into the experimental setup tomorrow and perform terribly using the exact same classifier. One initial approach to get over this problem was to fix the classification rule beforehand and train the patient to force brain activity to conform to this rule. For example, Wolpaw et al. in the early 90's chose weights for the $\alpha$ and $\mu$ rhythms and trained participants to modulate the bandpower in these frequency bands in order to control a cursor \cite{Wolpaw1991,Wolpaw1994}. Similarly, Birbaumer et al. trained a patient to create large depolarizations of the electroencephalogram (EEG) over the course of several seconds, using a simple threshold on the bandpassed raw signal \cite{Birbaumer1999}. In their time, both approaches were successful, but took training time on the order of months to master. To overcome this limitation, several groups introduced machine learning techniques for adapting BCIs to their users \cite{Ramoser2000, Blankertz2002, Lal2004, Grosse-Wentrup2007a, Hill2006, Blankertz2007, Grosse-Wentrup2008}. They successfully managed to learn decoding rules with high acccuracy using only a 
fraction of the training trials required by the earlier approaches, allowing subjects to communicate consistently with a computer in a single session. Unfortunately, a training period had to
be repeated at the beginning of each usage session as the learned discrimination rules were not immediately stable. A naive solution to this limitation was to pool training data from multiple recordings; however, the statistical distributions of these data varies across subjects as well as across sessions within individual subjects, giving this approach varying effectiveness. In recent years, several groups have started explicitly modelling such variations to exploit structure that is shared between data recorded from multiple subjects and/or sessions. In this article, we provide an overview of previous work on the topic and present a unifying approach to transfer learning in the field of BCIs. We demonstrate the utility of our framework on subject-to-subject transfer in a motor-imagery paradigm as well as on session-to-session transfer in one patient diagnosed with amyotrophic lateral sclerosis (ALS).

\subsection{Previous work}
\label{sec:review}
\begin{figure}
\centering
\includegraphics[width=0.7\textwidth]{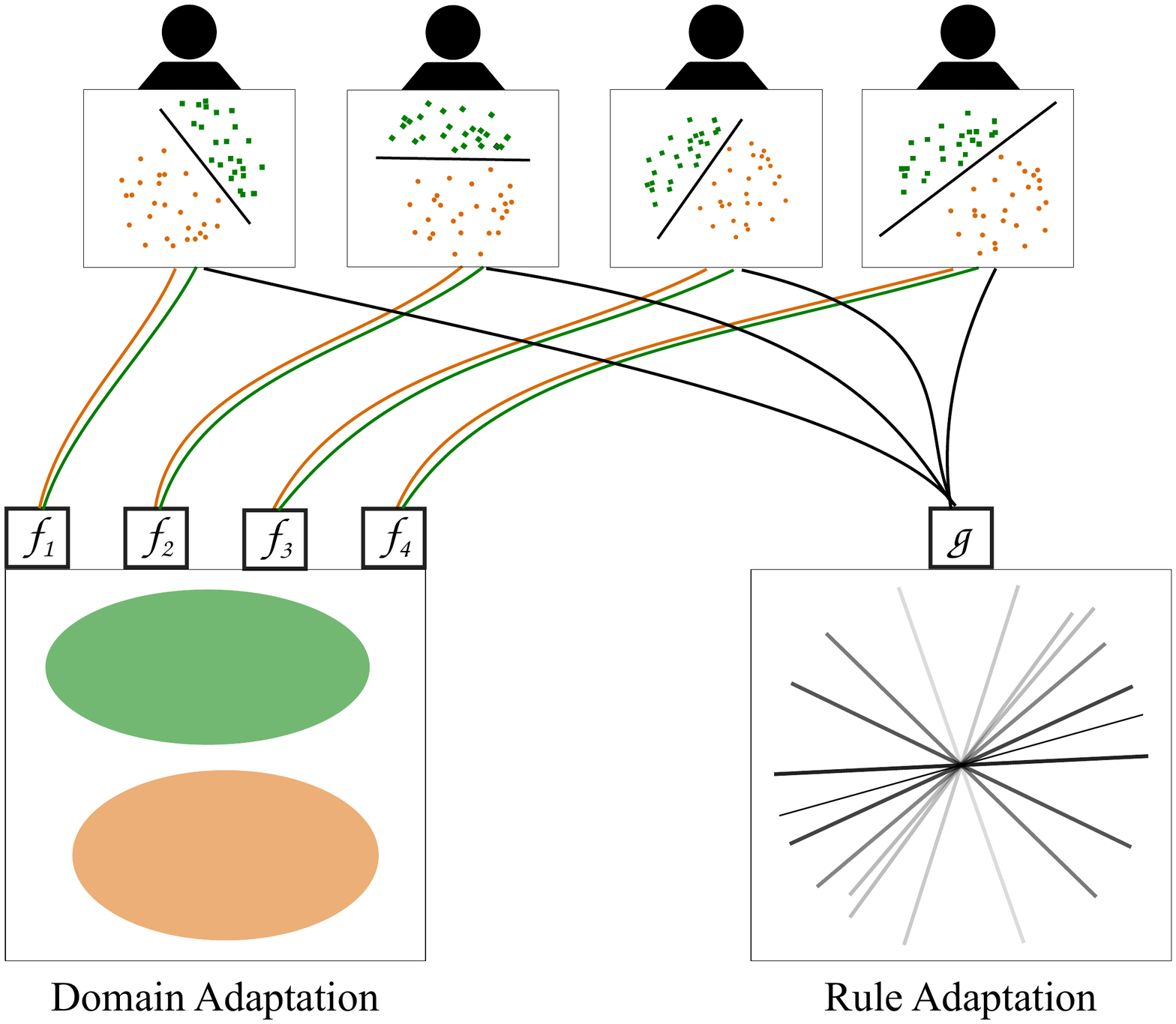}%
\caption{\label{Fig:TL}Given a set of training datasets (top) there are two ways to model the similarities shared by them. \emph{Domain adaptation} (left) refers to the strategy of attempting to find a transformation to a data space in which a single decision rule will classify all samples. Instead of learning a new rule for the new data, data is simply transformed to the invariant space. \emph{Rule adaptation} (right) is the strategy of attempting to learn the structure of the classification rules. New datasets are faced with a much smaller search space of possible rules which allows for much faster learning of novel decision boundaries.}
\end{figure}

\textit{Transfer learning} describes the procedure of using data recorded in one task to boost performance in another, related task (for a more exhausive review of the machine learning  literature, see \cite{Pan2010}). That is to say, we assume \emph{a priori} that there is some structure shared by these tasks; the goal, then, is to learn some representation of this structure so further tasks can be solved more easily. In the context of BCIs, transfer learning is of critical importance - it has long been known that the EEG signal is not stationary, and so in its strictest sense one can consider every trial a slightly new task. As such, long sessions of BCI usage present unique problems in terms of consistent classification \cite{Abbass2014}. The question is how to transfer some sort of knowledge between them: a question that can be answered in one of two general ways. Either we can attempt to find some structure in the data that is invariant across datasets or we can find some structure in how the decision rules differ between different subjects or sessions. We denote these as \emph{domain adaptation} and \emph{rule adaptation} respectively (Figure \ref{Fig:TL}).

Looking at the literature, BCI has been almost exclusively dominated by domain adaptation approaches. One popular feature space in the field is the trial covariance matrices used both in Common Spatial Patterns (CSP)\cite{Koles1991,Ramoser2000} and other more modern methods \cite{Barachant2010}. Many transfer learning techniques have been attempted with CSP, mostly relying on an assumption that there exists a set of linear filters that is invariant across either sessions or subjects. An early example of session-to-session transfer of spatial filters is the work by Krauledat et al.~\cite{Krauledat2008}, in which a clustering procedure is employed to select prototypical spatial filters and classifiers, which are in turn applied to newly recorded data. Using this approach, the authors demonstrate that calibration time can be greatly reduced with only a slight loss in classification accuracy. The problem of subject-to-subject transfer of spatial filters is addressed by Fazli et al.~\cite{Fazli2009}: also building upon CSP for spatial filtering, the authors utilize a large database of pairs of spatial filters and classifiers from 45 subjects to learn a sparse subset of these pairs that are predictive across subjects. Using a leave-one-subject-out cross-validation procedure, the authors then demonstrate that this sparse subset of spatial filters and classifiers can be applied to new subjects with only a moderate performance loss in comparison to subject-specific calibration. Note that in both above approaches transfer learning amounts to determining invariant spaces on which to project the data and learning classifiers in these spaces. This line of work has been further extended by Kang et al.~\cite{Kang2009,Kang2014}, Lotte and Guan \cite{Lotte2011}, and Devlaminck et al.~\cite{Devlaminck2011}. In these contributions, the authors demonstrate successful subject-to-subject transfer by regularizing spatial filters derived by CSP with data from other subjects, which amounts to attempting to find an invariant subspace on which to project the data of new subjects. Recently, a method of distance measures between trial covariance matrices has also been used to great effect in both motor imagery \cite{Barachant2012} and event-related potential paradigms \cite{Congedo2015} as a domain adaptation tool. Related to the spirit of the regularized CSP methods described above, they work by trying to find the best projection plane for the trial covariance matrices, invariant to subjects and sessions, and then run a classification algorithm. Other domain adaptation approaches include that by Morioka et al.~\cite{Morioka2015}, in which an invariant sparse representation of the data is learned using many subjects and then the transformation into that space is applied to new subjects, and the technique of stationary subspace analysis \cite{Buenau2009,Samek2012}, which attempts to find a stationary subspace of the data from multiple subjects and/or sessions. 

A very related technique to domain adaptation is \emph{covariate shift}, which has also found use in BCIs. Sugiyama et al.~have used covariate shift adaptation to combine labeled training data with unlabeled test data \cite{Sugiyama2007}. Here, it is assumed that the marginal distribution of the data changes between the subjects and/or sessions, but the decision rule with respect to this marginal distribution remains constant. This assumption leads to a re-weighting of training data from other subjects and/or previous sessions based on unlabeled data from the current test set that corrects for covariate shifts--in essence, correcting for the difference in marginal distributions in the different subjects and/or sessions. In addition to their results, several other authors have also reported improvements in BCI decoding performance by using similar techniques for covariate shift adaptation \cite{Li2010, Mohammadi2013, Arvaneh2014}. Other techniques such as boosting \cite{Dai2007} have also used re-labelling of offline data to increase performance \cite{Yang2014}. 

The covariate shift and other methods presented in the previous paragraph represent a very different assumption about the tasks than the methods that attempt to find an invariant space to project the data. Instead of assuming that there exists some space where the data already lives that is invariant for all individuals or across all time, it attempts to model the variation between individuals and efficiently discover a transformation for new individuals to the known space (in comparison, an invariant subspace could be seen as applying an identical transform to all individuals). This approach of attempting to learn a representation of the variability is most naturally attempted in the space of possible rules, since it often offers a ready-made parametrization of the approximating function. One possibility for such modelling is to treat the parameters of a decoding model as random variables that are, for each subject and/or session, drawn from the same distribution. The prior distribution of the model parameters can then be used to link training data across multiple subjects and/or sessions, and be learned by a simultaneous optimization over previous subjects and/or sessions. Rule adaptation of this sort has been attempted in Kinderman et al.~\cite{Kindermans2012}, which attempts to learn a classification prior in the P300 task, but restricts the covariance to multiples of the identity while it allows the mean to be determined by the distribution of subject weight vectors. A framework of \emph{multitask learning} which attempts to learn a full distribution has been introduced to the field of BCIs by Alamgir et al.~\cite{Alamgir2010}. Specifically, the authors treat classification as a linear regression problem and model the regression weights as a random variable that is drawn from a multivariate Gaussian distribution with unknown mean and covariance matrix. By jointly estimating the parameters of this distribution and regression weights for multiple subjects, they demonstrate a substantial improvement in decoding performance in a motor-imagery paradigm. However, this work suffered from various limitations. In modelling each channel bandpower as a separate feature it became necessary to employ channel selection in a pre-processing step, and also to
attempt to isolate and remove noisy subjects from the training pool. In this work we extend the previous results of multitask learning with a new technique that is robust both
to subjects who perform poorly and to an extremely high-dimensional feature space. 

%% file: Training.tex
\section{A general framework for transfer learning in BCIs}
\label{sec:offline}

In this article, we build upon our prior work on multitask learning \cite{Alamgir2010} to derive a general framework for transfer learning in BCI, applicable to any spatiotemporal feature space and able to be used on multi-session and multi-subject data equally, and further introduce a BCI-specific method for reducing the feature space dimension. 

\subsection{Preliminaries}

In this section, we introduce the decoding model used throughout this work. We index multiple subjects or recording sessions by $s=\{1,\dots,S\}$ and assume that for each subject/session we are given data from $n_s$ trials, $D_s = \{(\x^i_s,y^i_s)\}_{i=1}^{n_s}$.
Here, $\x^i_s \in \mathbb{R}^d$ refers to the features derived from the
recorded brain signals of subject/session $s$ during trial $i$, with $d$ denoting the
number of features. For the datasets presented in this
article, $\x^i_s$ consists of EEG log-bandpower estimates at
different scalp locations; however, it is equally applicable to timepoints after event onset if the signal of interest is an event-related potential. More specifically, if the number of electrodes
is $E$ and the number of EEG log-bandpower estimates is $F$, the number of
features is $d=E \times F$.
Variable $y^i_s$ denotes the subject's stimulus, e.g., motor imagery of either the left or right hand in trial $i$ of session $s$. As we furthermore only deal with two-class paradigms, we let $y^i_s \in \{-1,1\}$ for all $i$ and $s$, though this framework is applicable to regression problems as well.

Assuming our model is linear with a noise term $\eta$, we can model our data by a linear function 
\begin{eqnarray*}
y^i_s = \w_s^{\text{T}}\x_s^i + \eta
\end{eqnarray*}

\noindent associated to each subject/session $s$, where
the parameters $\w_s$ constitute the weights assigned to the
individual features that are used to predict the stimulus for trials in
subject/session $s$.
 Given a  new brain signal  $\x$ for subject/session $s$,
the stimulus is predicted by
\begin{equation}
\hat{y}^{i+1}_s = \text{sign}\{ \w_s^{\text{T}}\x_s^{i+1} \}.
\label{eq:predictor}
\end{equation}
We first investigate training $\w_s$ for each subject independently
in Section \ref{sec:ind} and extend this formalism to train $\w_s$ jointly
on multiple subjects/sessions in Section \ref{sec:joint}. 

\subsection{Training Models for Subjects/Sessions Independently}
\label{sec:ind}
When faced with some set of data and labels, the goal is to determine the parameters $\w_s$ that allow for the best prediction of the labels from the data. 
Mathematically speaking, for each subject/session $s$, the parameters $\w_s$ are determined such that
the number of errors in the dataset of subject/session $s$, $D_s$, is small. The choice of how to define 'errors' for a given set of predictions can drastically influence both the values of the final parameters and the ease with which they can be found; in machine learning, this is called a \emph{loss function} and by finding the minimum of this function we can recover the parameters that result in the lowest defined error. The most commonly used
loss functions to calculate errors are convex proxies such as log-loss, hinge loss or least squares loss \cite{Hastie2001};
in this paper, we use least squares loss, which we arrive at naturally with the assumption that
the error term $\eta$ is distributed as $\mathcal{N}(0,\sigma^2)$. 

To begin, let us consider a probabilistic interpretation of the problem. Using Bayes Rule, the probability of our parameters given our data decomposes as follows (note that we ignore the possible dependence of the prior $p(\w_s)$ on $\x_s^i$ or $\sigma^2$):
  \begin{equation}
    p(\w_s|y_s^i,\x_s^i,\sigma^2)\propto p(y_s^i|\w_s,\x_s^i,\sigma^2)p(\w_s).
  \end{equation}
With the model from the previous section and the assumption of Gaussian noise, $p(y_s^i|\w_s,\x_s^i,\sigma^2) \sim \mathcal{N}(\w_s^{\text{T}}\x^i_s,\sigma^2)$, and assuming our samples $\x_s^i$ are independent, we may derive the negative log likelihood as follows:  \begin{align}
p(y_s^1,...,y_s^{n_s}|x_s^1,...,x_s^{n_s},\w_s,\sigma^2)&=\prod_{i=1}^{n_s}\mathcal{N}(y_s^i;\w_s^{\text{T}}\x_s^i,\sigma^2) \\\label{eq:least-sq}
LL(\w_s;D_s,\sigma^2)& = \frac{1}{\sigma^2}\sum_{i=1}^{n_s} \left( y^i_s - \w_s^{\text{T}}\x^i_s \right)^2,
\end{align}
The negative log likelihood defines a convenient loss function as its value increases with the square of the difference between our prediction $\w_s^{\text{T}}\x_s^i$ and the true label $y_s^i$ for each data point. For notational convenience, we write the loss in matrix form
by defining the input matrix $\X=[\x^{1 \text{T}},\dots,\x^{n \text{T}}]^{\text{T}}$ and the output vector
$\y=[y^1,\dots,y^n]^{\text{T}}$. Then, the loss for subject/session $s$ is given by
$\| \X_s \w_s - \y_s\|^2$, where $\|\mathbf{v}\|$ is the $\ell^2$ or Euclidean norm. If we ignore the prior and solve for $\w_s$ analytically from here, we end up with the equations for regular linear regression.

It is well known that complex models
that are trained without a validation dataset can {\it over-fit}, leading to poor
generalization to new data points. A classical technique to control over-fitting
is adding a penalty term to the loss function that reduces the complexity
of the model. A common choice for this regularizer is given by the sum
of the squares of the weight parameters,
\begin{equation}
\Omega(\w_s) = \frac{\| \w_s \|^2}{2}.
\label{eq:RR}
\end{equation}
Addition of $\Omega(\w_s)$ to the optimization problem is equivalent to 
 assuming a Gaussian prior on $\w_s$ with $\0$ mean and  unit covariance
$\I$ and incorporating this prior in the log-scale.\footnote{Note that $LL(\w_s;D_s,\sigma^2)+
\Omega(\w_s)$ gives the negative log posterior for $\w_s$ given $D_s$ and the assumed prior.} If the variance of the prior is not assumed to be exactly the identity matrix but rather some matrix $\alpha\I$ then this formulation describes ridge regression. 

However, the above assumption is rarely a reasonable one. If there exists some better prior information on the distribution of the weights that can be represented by a mean $\mmu$ and covariance $\SSigma$, this information can be used instead in the regularizer by
assuming a Gaussian distribution with the corresponding mean and covariance
term, $\Ncal(\mmu,\SSigma)$, as the prior and defining the
regularizer as the negative log prior probability
\begin{equation}
  \Omega(\w_s;\mmu,\SSigma) = \frac{1}{2} \left[ (\w_s - \mmu)^{\text{T}} \SSigma^{-1} (\w_s - \mmu)
\right] 
 + \frac{1}{2} \log \det(\SSigma).
\label{eq:prior}
\end{equation}
Note that the last term is constant with respect to $\w_s$ for
fixed $\SSigma$, and further that $\Omega(\w_s; \mathbf{0},\I)$ is equivalent to (\ref{eq:RR}).

The new loss function can then be derived by taking the negative logarithm of the posterior of $y_s$:
\normalsize
\begin{equation}
p(y^i_s|\w_s,\x_s^i,\mmu,\SSigma,\lambda) \propto \mathcal{N}(y^i_s;\w_s^{\text{T}}\x^i,\lambda)\mathcal{N}(\w_s;\mmu,\SSigma)
\end{equation}
\begin{equation}\label{eq:loss-ind}
LP(\w_s;D_s,\mmu,\SSigma,\lambda) = \frac{1}{\lambda} \| \X_s \w_s - \y_s\|^2 
 + \Omega(\w_s;\mmu,\SSigma) + \text{C}.
\end{equation}
We replace $\sigma^2$ with $\lambda$ to emphasize that in the loss function, the variance of the original noise model is equivalent to a term that controls the ratio of the importance assigned to the prior probability of the learned weight vector versus how well the learned vector can predict the labels in the training data. Put another way, the higher the variance of the noise in the model, the less we can trust our training data to lead us to a good solution; moving forwards, it is more convenient to think of the variable in terms of this trade-off than as purely a noise variance. From this point the actual optimization problem can be formulated as 
\begin{equation}
    \min_{\w_s}LP(\w_s;D_s,\mmu,\SSigma,\lambda).
\end{equation}

\subsection{Training Models for Subjects/Sessions Jointly}
\label{sec:joint}

In a standard machine learning setting, there is a single prediction
problem or {\it task} to model and  there is usually no
prior information  on the distribution of the model parameters $\w$.
However, if there are multiple prediction tasks that are related to each other,
it is possible to use information from all the tasks in order to improve
the inferred model of each task. In particular,
if the tasks share a common structure along with some task-specific variations,
the shared structure can be used as the prior information $(\mmu,\SSigma)$ in
(\ref{eq:prior}) in order to ensure that the solutions to all the tasks are close to each other in some space. 

In the BCI training problem, we treat each subject/session as one task and
the shared structure $(\mmu,\SSigma)$
represents the subject/session-invariant characteristics of stimulus prediction.
More precisely, $(\mmu,\SSigma)$ are
the mean vector and  covariance matrix  of features.
As such, $\mmu$ defines an \ob BCI that can be used to classify data recorded
from a novel subject/session without any subject/session-specific calibration process.
The divergence of a subject/session model
from the shared structure, $\|\w_s - \mmu\|$, represents the subject/session-specific
characteristics of the stimulus prediction.

Clearly, the shared structure is unknown in this setting.
Our goal is to infer the shared structure, $(\mmu,\SSigma)$, from
all the tasks along with the model parameters $\w_s$ jointly.
This can be achieved by combining the optimization problem of all tasks
\begin{equation}\label{eq:loss}
\min_{\W,\mmu,\SSigma} LP(\W,\mmu,\SSigma;D,\lambda) =
\min_{\W,\mmu,\SSigma} \frac{1}{\lambda} \sum_s \| \X_s \w_s - \y_s\|^2
 + \sum_s\Omega(\w_s;\mmu,\SSigma)
\end{equation}
where $\W=[\w_1,...,\w_S]^{\text{T}}$, $D=\{D_s\}_{s=1}^S$, and $d$ is the dimension of each weight vector.
Let us investigate each term of this optimization problem separately.
The first term is the sum of the losses from each session, and by minimizing it we ensure all
the sessions are well fitted.
The second term controls the divergence of
each subject/session model from the underlying mean vector $\mmu$ and penalizes
the elements of the residual $\hat{\w}_s=\w_s-\mmu$ scaling with $\SSigma^{-1}$.
Expanding  one of these terms,
\begin{equation*}
\hat{\w}_s^{\text{T}} \SSigma^{-1} \hat{\w}_s =  \sum_i{\sum_j \SSigma^{-1}_{i,j} \hat{\w}_{s,i} \hat{\w}_{s,j}},
\label{eq:penalize}
\end{equation*}
we observe that
$\SSigma^{-1}_{i,j}$ is proportional to the partial correlation between the $i$-th and $j$-th components of the weight vector,
which is defined as the correlation between these after all other components have been regressed out. Thus, for a given matrix $\SSigma^{-1}$, this term will be minimized when for each set of components with high partial correlation,
the subject/session-specific weight vectors $\w_s$ allow only one of these to deviate greatly from the mean of that component. Hence, $\SSigma^{-1}$ acts as an implicit feature selector.
The final term, which is a constant in the independent setting of
(\ref{eq:loss-ind}), controls the complexity of the covariance matrix.

We solve the minimization in (\ref{eq:loss}) with respect to $\W$ and $(\mmu,\SSigma)$
iteratively by alternating holding $(\mmu,\SSigma)$ and $\W$ constant.
For fixed $\mmu$ and $\SSigma$, optimization over
$\w_s$ decouples across subjects/sessions and hence can be optimized independently.
In each iteration we get the new $\w_s$ by
taking the derivative with respect to $\w_s$ for all $s$ and equating to 0.
This yields the following closed form update for each $\w_s$:
\begin{equation}
\w_s = \left(\frac{1}{\lambda}\X^{\text{T}}_s \X_s + \SSigma^{-1}\right)^{-1}
  \left(\frac{1}{\lambda} \X_s^{\text{T}}\y_s + \SSigma^{-1} \mmu\right).
\label{eq:w-update1}
\end{equation}
Hence, the model parameters are a combination of the shared model
contribution $\SSigma^{-1} \mmu$ and the contribution of the individual
subject/session data $\X_s^{\text{T}}\y_s$. This combination is scaled with
the inverse of covariance term which again comes
from both the data $\X^{\text{T}}_s \X_s$
and the shared model $\SSigma^{-1}$.
In order to avoid inverting $\SSigma$, which is a $O(d^3)$ operation,
we perform the equivalent update
\begin{equation}
\w_s = \left(\frac{1}{\lambda}\SSigma \X^{\text{T}}_s \X_s  + \I\right)^{-1}
  \left(\frac{1}{\lambda} \SSigma \X_s^{\text{T}}\y_s + \mmu \right).
\label{eq:w-update}
\end{equation}

For fixed $\W$, the updates of $\mmu$ and $\SSigma$ are given in Algorithm \ref{alg:algo} and derived in the Supplementary Materials.

\begin{algorithm}[tbh]
\caption{\label{alg:algo}
Multitask BCI training}
\begin{algorithmic}[1]
  \STATE {\bfseries Input:} $D$, $\lambda$
\STATE Set $\{(\mmu,\SSigma)\} = (\0,\I)$
\REPEAT
\STATE Update $\w_s$ using (\ref{eq:w-update})
\STATE Update $\mmu$ using $\mmu^* = \frac{1}{S} \sum_s \w_s$
\STATE Update $\SSigma$ using $\SSigma^* = \frac{ \sum_s (\w_s - \mmu) (\w_s - \mmu)^{\text{T}}}
{\tr \left( \sum_s (\w_s - \mmu) (\w_s - \mathbf{\mmu})^{\text{T}} \right)  }+\epsilon \I$
\UNTIL convergence
  \STATE {\bfseries Output:} $(\mmu,\SSigma)$
\end{algorithmic}
\end{algorithm}

\subsection{Decomposition of Spatial and Spectral Features}
\label{sec:decomp}

The learning method described above can be
applied to any feature representation where the features extracted from each electrode are appended together.
Let $E$ be the number of electrodes and $F$ be the number of
features extracted from each electrode. The final feature vector then is
size $EF$, rendering the covariance matrix large and iterative
updates expensive. It also causes the number of features to grow
linearly with the number of channels and channel-specific features,
an increase that can be avoided by taking advantage of the structure
of the EEG. Specifically, we assume that the contribution
of the features is invariant across electrodes
but the importance of each electrode varies. Hence, the weights corresponding to the feature vector mentioned above can be
decomposed into two components: the weight
of each electrode  $\aalpha=(\alpha_1, \dots, \alpha_E)$
and the weights of features that are shared across all electrodes
$\w=(w_1, \dots, w_{F})$. We note that though in this paper spectral features are used, there is no reason that
temporal features such as ERP timepoints could not be used instead. With this formulation, the linear regression function is given
by
\begin{equation*}
f_s(X;\w_s,\aalpha_s) = \aalpha_s^{\text{T}}X\w_s,
\end{equation*}

\noindent where $X \in \mathbb{R}^{E \times F}$ denotes the matrix of features for each channel for a given trial. This causes
the number of parameters in the decoding model to be reduced from $EF$ to $E + F$.

The new optimization problem is now over $\W,\A,\mmu_{\w},\mmu_{\aalpha},\SSigma_{\w},$ and $\SSigma_{\aalpha}$, where $\A=[\aalpha_1,...,\aalpha_S]^{\text{T}}$.
However, it can easily be seen that $\aalpha^{\text{T}}X\w=\aalpha^{\text{T}} \tilde X$, where $\tilde X
= X\w$, and thus that $y$, instead of being a function of the features, can now be considered a function
of the aggregated features for each electrode. As this formulation assumes that $\aalpha$ and $\w$ are indepedent, the prior over model
parameters can be incorporated as the product of indepedent priors for both $\w$ and $\aalpha$. As such, the same arguments used to 
define a prior of $\w$ can be applied to $\aalpha$ to define a new distribution for $y_s^i$ and a new
optimization problem (for readability we define the parameters of the Gaussian priors over $\w$ and $\aalpha$ as
$\theta_\w$ and $\theta_{\aalpha}$ respectively):
\begin{equation}
p(y^i_s|X_s^i,\w_s,\aalpha_s,\theta_\w,\theta_{\aalpha},\lambda) \propto 
\mathcal{N}(y^i_s;\aalpha^{\text{T}}_sX_s^i\w_s,\lambda)\mathcal{N}(\w_s;\theta_{\w})\mathcal{N}(\aalpha_s;\theta_{\aalpha}) 
\end{equation}
\begin{equation}\label{eq:loss-decomp}
LP(\W,\A,\theta_\w,\theta_{\aalpha}|D,\lambda) = \frac{1}{\lambda} \sum_s \sum_i\|
\aalpha_s^{\text{T}}X_s^i\w_s - y_s^i\|^2 
+\sum_s \Omega(\w_s;\mmu_\w,\SSigma_\w) + \sum_s \Omega(\aalpha_s;\mmu_{\aalpha},\SSigma_{\aalpha}) + \text{C}
\end{equation}
where again, $\Omega(\x;\mmu,\SSigma)$ is the negative log prior probability of the vector $\x$ given the Gaussian distribution parametrized by $(\mmu,\SSigma)$. It is easy to see that $\w$ and $\aalpha$ function identically except for a  
transpose. The updates for the weights over the features and the channels are linked, so we first iterate until 
convergence within each subject/session before continuing on to update the prior parameters, which leads to the following 
algorithm (Algorithm \ref{alg:FDalgo}):

\begin{algorithm}[tbh]
\caption{\label{alg:FDalgo}
Multitask BCI training with uninformative $\aalpha$ }
\begin{algorithmic}[1]
  \STATE {\bfseries Input:} $D$, $\lambda$
\STATE Set $\{(\mmu_{\w},\SSigma_{\w}),(\mmu_{\aalpha},\SSigma_{\aalpha})\} = (\0,\I)$
\STATE Set $\aalpha_s=\textbf{1}$
\REPEAT
\REPEAT
\STATE Compute $\tilde{\X}_s= [\aalpha_s^{\text{T}}X_s^1;\dots;\aalpha_s^{\text{T}}X_s^n]$
\STATE Compute $\widehat{\X}_s = [X_s^1\w_s,\dots,X_s^n\w_s]$
\STATE Update $\w_s$ using $\w_s^* = \left(\frac{1}{\lambda}\SSigma_\w \tilde{\X}_s^{\text{T}} \tilde{\X}_s  + \I\right)^{-1}
  \left(\frac{1}{\lambda} \SSigma_\w \tilde{\X}_s^{\text{T}}\y_s + \mmu_\w \right)$
\STATE Update $\aalpha_s$ using  $\aalpha_s^* = \left(\frac{1}{\lambda}\SSigma_{\aalpha} \widehat{\X}_s \widehat{\X}_s^{\text{T}}  + \I\right)^{-1}
  \left(\frac{1}{\lambda} \SSigma_{\aalpha} \widehat{\X}_s\y_s + \mmu_{\aalpha} \right)$
\UNTIL $\W$ and $\mathbf{A}$ converge for fixed $(\mmu,\SSigma)$
\STATE Update $\mmu_\w,\mmu_{\aalpha}$ using $\mmu_w^* = \frac{1}{S} \sum_s \w_s, \mmu_{\aalpha}^* = \frac{1}{S} \sum_s \aalpha_{s}$
\STATE Update $\SSigma_\w,\SSigma_{\aalpha}$ using $\SSigma_\w^* = \frac{ \sum_s (\w_s - \mmu_\w) (\w_s - \mmu_\w)^{\text{T}}}
{\tr\left( \sum_s (\w_s - \mmu_\w) (\w_s - \mmu_\w)^{\text{T}}\right) }+\epsilon \I, 
\SSigma_{\aalpha}^* = \frac{ \sum_s (\aalpha_s - \mmu_{\aalpha}) (\aalpha_s - \mmu_{\aalpha})^{\text{T}}}
{\tr\left( \sum_s (\aalpha_s - \mmu_{\aalpha}) (\aalpha_s - \mmu_{\aalpha})^{\text{T}}\right) }+\epsilon \I$
\UNTIL convergence
  \STATE {\bfseries Output:} $(\mmu_\w,\SSigma_\w,\mmu_{\aalpha},\SSigma_{\aalpha})$
\end{algorithmic}
\end{algorithm}

This reduces the size of the feature space from $EF$ to $E+F$, which simplifies learning the
regression parameters and also reduces calculation speed. The more degrees of freedom, the more data a model requires
to find a good fit, so by reducing the number of parameters we also reduce the number of necessary training trials.
Also for the case of a model with $EF$ parameters, the 
matrix inversion necessary to compute a decision rule is $O(E^3F^3)$, which is changed for a model with $E+F$ parameters
to $O((E+F)^3)$. We also note that the initialization of Algorithm \ref{alg:FDalgo} shown above is non-informative. Our experiments have suggested that
the alternative method shown below (Algorithm 3) works more effectively in some cases.

\begin{algorithm}[tbh]
\caption{\label{alg:FDalgo_pooledA}
Multitask BCI training with $\aalpha$ initialization }
\begin{algorithmic}[1]
\STATE {\bfseries Input:} $D$, $\lambda$
\STATE Set $\{(\mmu_\w,\SSigma_\w),(\mmu_{\aalpha},\SSigma_{\aalpha})\} = (\0,\I)$
\STATE Concatenate subject data in $D$ into single pooled subject $\hat D$
\STATE Run ridge regression on $\hat D$ using the feature decomposition regression function
\STATE Set $\aalpha_s$ to the ridge regression spatial weights
\REPEAT
\REPEAT
\STATE Compute $\tilde{\X}_s= [\aalpha_s^{\text{T}}X_s^1;\dots;\aalpha_s^{\text{T}}X_s^n]$
\STATE Compute $\widehat{\X}_s = [X_s^1\w_s,\dots,X_s^n\w_s]$
\STATE Update $\w_s$ using $\w_s = \left(\frac{1}{\lambda}\SSigma_\w \tilde{\X}_s^{\text{T}} \tilde{\X}_s  + \I\right)^{-1}
  \left(\frac{1}{\lambda} \SSigma_\w \tilde{\X}_s^{\text{T}}\y_s + \mmu_\w \right)$
\STATE Update $\aalpha_s$ using  $\aalpha_s = \left(\frac{1}{\lambda}\SSigma_{\aalpha} \widehat{\X}_s \widehat{\X}_s^{\text{T}}  + \I\right)^{-1}
  \left(\frac{1}{\lambda} \SSigma_{\aalpha} \widehat{\X}_s\y_s + \mmu_{\aalpha} \right)$
\UNTIL $\W$ and $\mathbf{A}$ converge for fixed $(\mmu,\SSigma)$
\STATE Update $\mmu_\w,\mmu_{\aalpha}$ using $\mmu_w^* = \frac{1}{S} \sum_s \w_s, \mmu_{\aalpha}^* = \frac{1}{S} \sum_s \aalpha_{s}$
\STATE Update $\SSigma_\w,\SSigma_{\aalpha}$ using $\SSigma_\w^* = \frac{ \sum_s (\w_s - \mmu_\w) (\w_s - \mmu_\w)^{\text{T}}}
{\tr\left( \sum_s (\w_s - \mmu_\w) (\w_s - \mmu_\w)^{\text{T}}\right) }+\epsilon \I, 
\SSigma_{\aalpha}^* = \frac{ \sum_s (\aalpha_s - \mmu_{\aalpha}) (\aalpha_s - \mmu_{\aalpha})^{\text{T}}}
{\tr\left( \sum_s (\aalpha_s - \mmu_{\aalpha}) (\aalpha_s - \mmu_{\aalpha})^{\text{T}}\right) }+\epsilon \I$
\UNTIL convergence
  \STATE {\bfseries Output:} $(\mmu_\w,\SSigma_\w,\mmu_{\aalpha},\SSigma_{\aalpha})$
\end{algorithmic}
\end{algorithm}

\subsection*{Online resource for multitask learning}
Supplementary materials, appendix, and MATLAB and Python implementations of all three algorithms described here can be found at 
http://brain-computer-interfaces.net/.

%% file: Online.tex
\subsection{Adaptation to Novel Subjects}
\label{sec:online}

In Section \ref{sec:offline}, we  outlined a simple yet effective approach
to infer the subject-invariant BCI model, given by learning the parameters of a Gaussian distribution over the weights.
This model can be used successfully on novel subjects immediately
via $f(\x;\theta)=\mmu^{\text{T}}\x$ in the case of regular linear regression or $f(X;\theta_\w,\theta_{\aalpha}) = \mmu_{\aalpha}^{\text{T}}X\mmu_\w$
in the case of feature decomposition, 
though depending on the covariance of the learned priors this can result in poor performance. It is possible to further improve the performance of this model
by adapting to the subject as more subject-specific data becomes available by simply using the learned priors and considering the problem independently as
discussed in Section \ref{sec:ind}. The standard regression case is discussed there; for the feature decomposition method, we consider $n$ trials $X^i$,
where each $X^i \in \mathbb{R}^{E \times F}$ is a matrix with columns denoting features and rows denoting electrodes. In this setting the update equations
are identical to the inner loop of Algorithm \ref{alg:FDalgo}. We emphasize that $\w$ and $\aalpha$ are linked, so the update steps must be iterated until convergence. The parameter $\lambda$ is determined in practice through
cross-validation over the training data.

%% file: MotorImagery.tex
\subsection{Subject-to-Subject Transfer}

\subsubsection*{Paradigm}
As an initial test of this algorithm, we considered how it performs on the most common paradigm in spectral BCIs: motor imagery. Specifically, subjects were placed in front of a screen with a centrally displayed fixation cross. Each trial started with a pause of three seconds. A centrally displayed arrow then instructed subjects to initiate haptic motor imagery of either the left or right hand, as indicated by the arrow's direction. After a further seven seconds the arrow was removed from the screen, marking the end of the trial and informing subjects to cease motor imagery.

\subsubsection*{Dataset}
Ten healthy subjects participated in the study (two females, $25.6 \pm 2.5$ years old). One subject had already participated twice in other motor imagery
experiments while all others were na\"{i}ve to motor imagery and BCIs. EEG data was recorded from 128 channels, placed according to the extended
10-20 system with electrode Cz as reference, and sampled at 500Hz.
BrainAmp amplifiers (BrainProducts, Munich) with a temporal analog
high-pass filter with a time constant of 10s were used for this purpose.
A total of $150$ trials per class (left/right hand motor imagery) per subject
were recorded in pseudorandomized order, with no feedback provided to the subjects during the experiment.

\subsubsection*{Feature Extraction}
For feature extraction, recorded EEG data was first spatially filtered using
a surface Laplacian setup \cite{McFarland1997}. We did not employ more
sophisticated methods for spatial filtering, such as CSP or beamforming,
in order to keep the spatial filtering setup data-independent.
For each subject, trial and electrode, frequency bands of $2$ Hz width,
ranging from 7-29 Hz, were then extracted using a discrete Fourier transform with a Hanning window,
computed over the length of the trial. Log-bandpower within the last seven seconds of each
trial for each frequency band then formed the $(128 \times 12)$-dimensional feature vector.

\subsubsection*{Classification Performance}
\label{subsec:classperf}
Here we show the efficiency of the proposed algorithms by examining the effect of multitask learning and FD on classification performance. For all algorithms, one subject was successively chosen as the test subject and all other subjects were then used for training. Test-specific training data
of between 10 and 100 trials per condition were then given to each algorithm, and the remaining trials out of 300 were used for testing. Multitask learning was done using Algorithm \ref{alg:algo} and Algorithm \ref{alg:FDalgo_pooledA} with a cross-validated $\lambda$. Note that for all tested algorithms the feature space was the full 128 channels, each with 12 feature bands.

We looked at two control algorithms to compare with the multitask learning approaches. The first was to consider ridge regression, which regularizes the
regression method only by penalizing the magnitude of the resultant weight vector (see (\ref{eq:RR})) and can be seen as using an uninformed prior for
the distribution of weight vectors; the second was to consider a support vector machine (SVM) trained on the same feature space. We further tested both control algorithms
two ways: Once with pooled data and once with only subject-specific data. For the pooled condition, all data from the training subjects
was concatenated to the training trials from the test subject to form a combined training set, on which the control algorithms were run. For the 
subject-specific condition, only training data from the test subject was used to train the control algorithms. All controls were compared to
the multitask approaches, where the learned prior mean(s) and covariance(s) were used to regularize the least-squares regression method. 

 The following list summarizes the algorithms:
 \begin{itemize}
     \item MT\_FD: multitask learning with Algorithm \ref{alg:FDalgo_pooledA} 
     \item MT: multitask learning with Algorithm \ref{alg:algo}
     \item RR: standard ridge regression
     \item RR\_FD: ridge regression using the FD regression method
     \item SVM: SVM with a linear kernel given the full $128 \times 12$ feature space
 \end{itemize}

\subsubsection*{Results}
\begin{figure}
\centering
\includegraphics[width=0.7\textwidth]{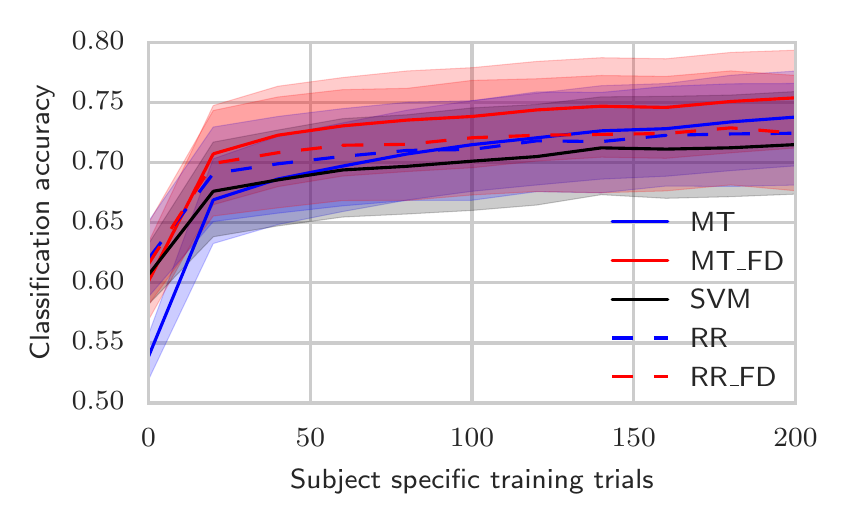}%
\caption{\label{Fig:ind_pooled_avg}Mean and STD (shaded) for classification accuracy of MT and pooled conditions across the ten subjects. The control algorithms were trained on data pooled across training subjects, and are compared against classification using Algorithm \ref{alg:algo} (MT, solid blue) and Algorithm \ref{alg:FDalgo_pooledA} (MT\_FD, solid red). Displayed
control algorithms are ridge regression using the standard regression method (RR, dashed blue), ridge regression using the FD regression method (RR\_FD, dashed red)  and SVM (SVM, solid black). The FD formulation of the multitask learning has
comparable performance with few training trials to pooled regression and both multitask algorithms manage to improve more than the pooled controls 
given a larger number of training trials. }
\end{figure}

\begin{figure} 
\centering
\includegraphics[width=0.7\textwidth]{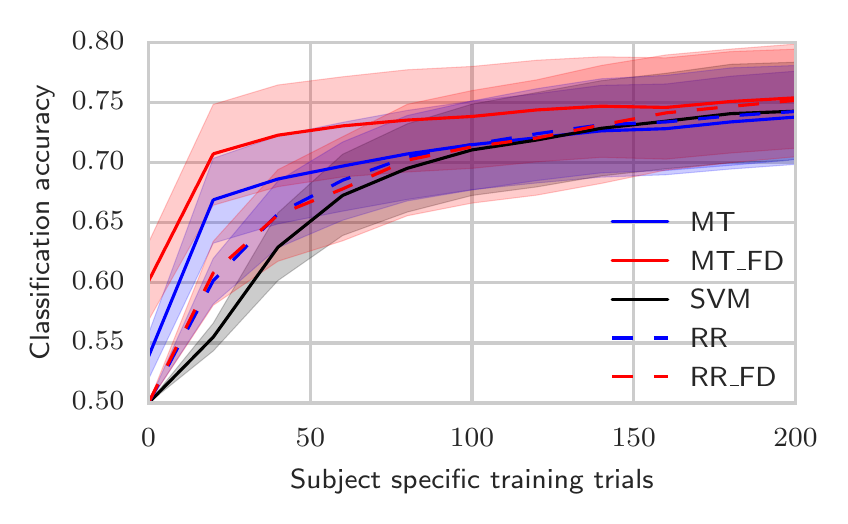}%
\caption{\label{Fig:ind_ss_avg}Mean and STD (shaded) for classification accuracy of MT and single-subject conditions across the ten subjects. Classification values for the multitask algorithms are identical to those shown in Figure \ref{Fig:ind_pooled_avg}. The control algorithms were trained on data exclusively from the test subject, and are compared against classification using Algorithm \ref{alg:algo} (MT, solid blue) and Algorithm \ref{alg:FDalgo_pooledA} (MT\_FD, solid red). Displayed
control algorithms are ridge regression using the standard regression method (RR, dashed blue), ridge regression using the FD regression method (RR\_FD, dashed red)  and SVM (SVM, solid black). The multitask algorithm with FD regression estimation performs better on average regardless of the number of trials, though single-subject ridge regression with the FD regression method manages to equal its performance at 200 training trials. }
\end{figure}

The results for the pooled sub-condition can be found in Figure \ref{Fig:ind_pooled_avg} and the results for the single-subject sub-condition can be found
in Figure \ref{Fig:ind_ss_avg}. Note that in both graph, the curves for the MT and MT\_FD algorithms are identical.

The MT\_FD algorithm consistently outperformed the other algorithms at nearly all levels of test subject data. In the pooled condition, it equalled the zero-training and low-data accuracy of the pooled data while also managing to more effectively use subject-specific data, leading to a higher mean accuracy than any other algorithms with 200 training trials. Interestingly, the MT algorithm without FD did more poorly than the
pooled ridge regression without FD in the zero-training and low subject-specific trial cases, which we hypothesize is due to the fact that each individual subject had so few training trials compared to the size of the feature space (300 compared to 1400). Rule adaptation requires learning a rule for each subject, which is hampered by this low number. However, by concatenating the trials together in ridge regression, pooling manages to work better. Regardless of this, MT without FD is still able to more effectively
use subject-specific data than any of the pooling algorithms as shown by the higher slope of the classification curve. The single-subject condition was used
to determine whether this regularization could reduce the maximum accuracy: With large training data and no data from different subjects, the best subject-specific rule can be found, and so we consider the maximum single-subject accuracy as an approximation of the maximum achievable accuracy with a linear boundary. We find that the MT
approaches at high numbers of trials achieve accuracies nearly identical to those achieved by only subject-specific training, showing that there is no reduction in maximum achievable accuracy for the MT approach. For subject-specific results please consult the Supplementary Materials.

\begin{figure}[t]
\subfloat{\includegraphics[trim=10 0 0 0,clip,width=4in]{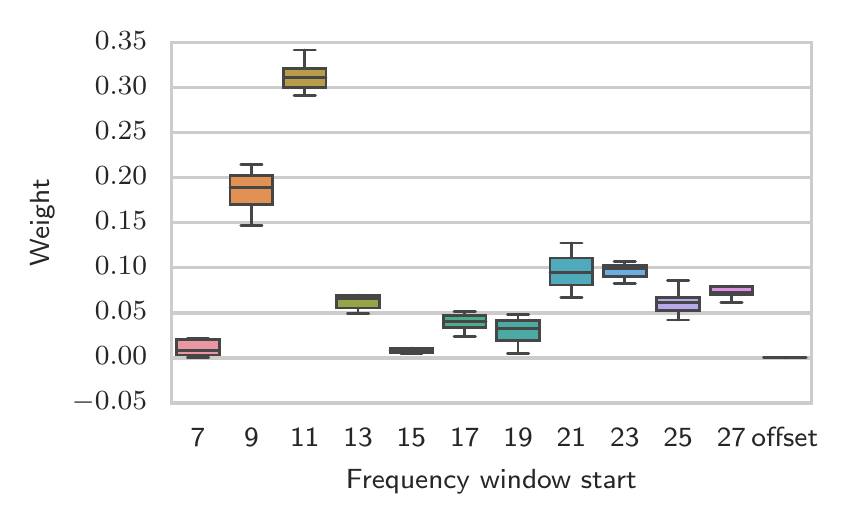}%
\label{Fig:MT_weight}}
\hfil
\subfloat{\includegraphics[trim=50 0 10 10,clip,width=3in]{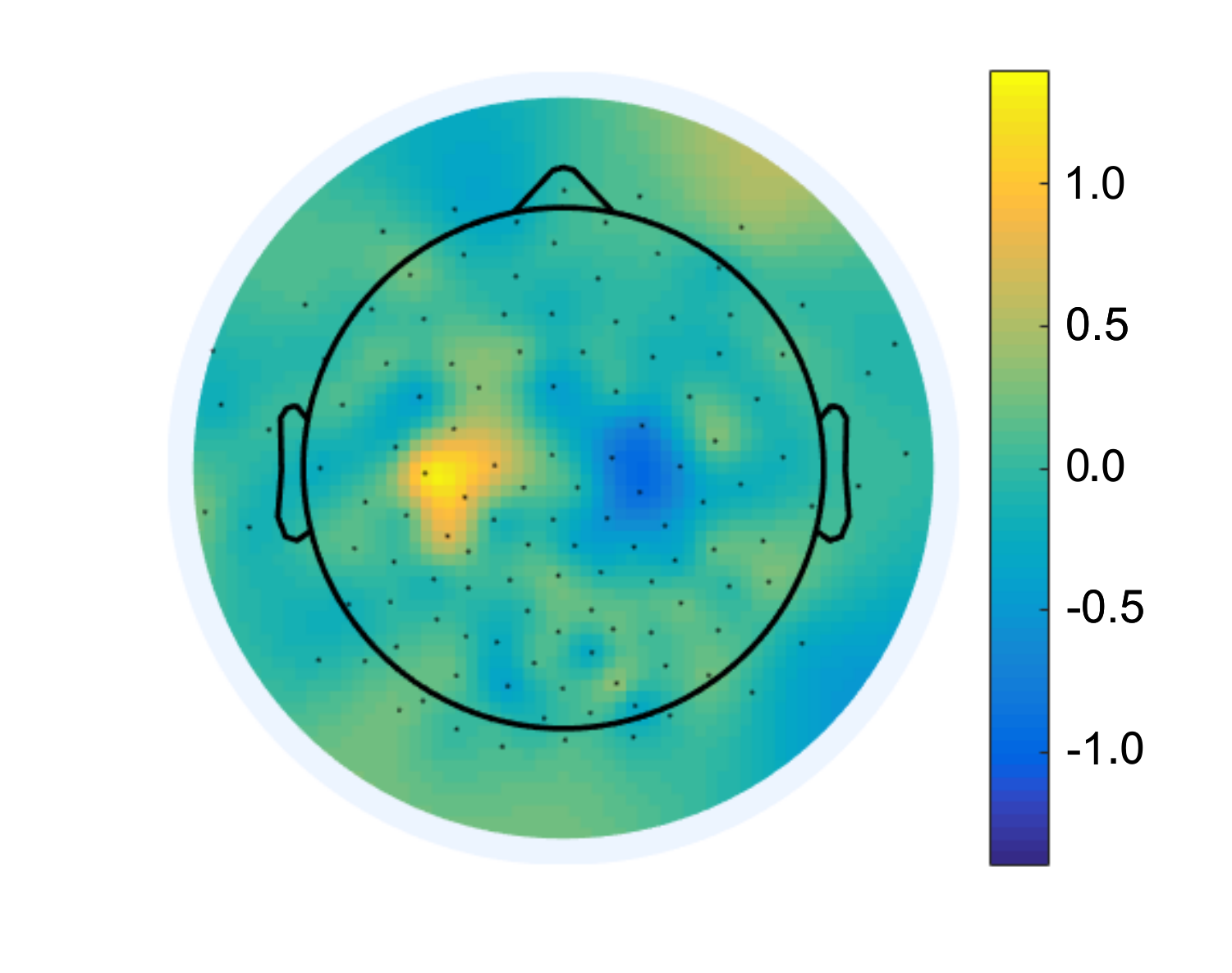}%
\label{Fig:MT_alpha}}
\caption{\label{Fig:MT_mus}(left) Box and whiskers plot of the absolute value of the learned prior means for all 10 leave-one-out executions of the MT algorithm, showing medians, first quartiles, and 1.5 times the IQR for the learned weights in each frequency range. The sign of the prior frequency mean is exchangeable with the sign of the spatial mean,
and so the absolute value is used here to correct for that. X-axis shows the starting frequency of each 2 Hz window. Note that the highest weights are concentrated around the windows of the $\mu$ frequency range. (right) Sum of the learned spatial weights over training subject groups
plotted topographically with respect to the head showing a concentration of high-magnitude weights over the motor cortices.}
\end{figure}

 To further confirm that our results are classifying on the signal we expect, we considered the mean of the spatial and spectral priors in the MT\_FD
 condition (Figure \ref{Fig:MT_mus}). 
 The learned topography is most strongly weighted around the electrodes directly over the motor cortices and the different
 cortices also have opposite signs, which is in agreement with spatial filters learned in CSP \cite{Koles1991,Ramoser2000} and beamforming \cite{Grosse-Wentrup2008}. Further, looking at the spectral weights, we see that the most important weight is on the $\mu$ band, which is consistent with 
 previous results on the subject, suggesting that our classification accuracy is indeed due to training on a brain-derived signal and not any sort of 
 artifact.


%% file: DeltaFeedback.tex
\subsection{Session-to-Session Transfer}
\label{sec:deltafeedback}

A common issue in BCI paradigms, especially those used with patient populations, is the low number of trials per session. Given the success of our
FD approach on the motor imagery data, we attempted it here on a 30-session dataset where each session had only ten training and between ten and twenty test trials for each condition.

\subsubsection*{Data Collection}
We trained a patient diagnosed with ALS to modulate the $\delta$-bandpower (1--5 Hz) in the precuneus in thirty sessions
over the course of fifteen months. The patient's ALS-FRS-R \cite{Cedarbaum1999} score decreased from 33 to 9 over the course of this time,
an average of 1.6 points per month. The paradigm setup is identical
to the setup in \cite{Grosse-Wentrup2014} except that the frequency band that received
feedback was 1--5 Hz and the target area was changed from the superior parietal cortex to the precuneus. In brief, however: The subject learned through operant conditioning
to modulate power in a beamformed signal pointed at the precuneus over the course of
sixty seconds,  deviating either up or down from a session-specific mean. Each minute-long interval
was one trial, and each run was twenty trials (ten up and ten down). For each session, the subject completed
between two and three runs. The first session was used entirely for training.

Throughout all sessions a 121-channel EEG was recorded at a sampling frequency
of 500 Hz using actiCAP active electrodes and a QuickAmp amplifier (both provided
by BrainProducts GmbH, Gilching, Germany).  Electrodes were placed according to the
extended 10-20 system with electrode Cz as the initial reference.  All recordings were
converted to common average reference. 

\subsubsection*{Feature extraction and training}
\begin{figure}
\centering
\includegraphics[width=5in]{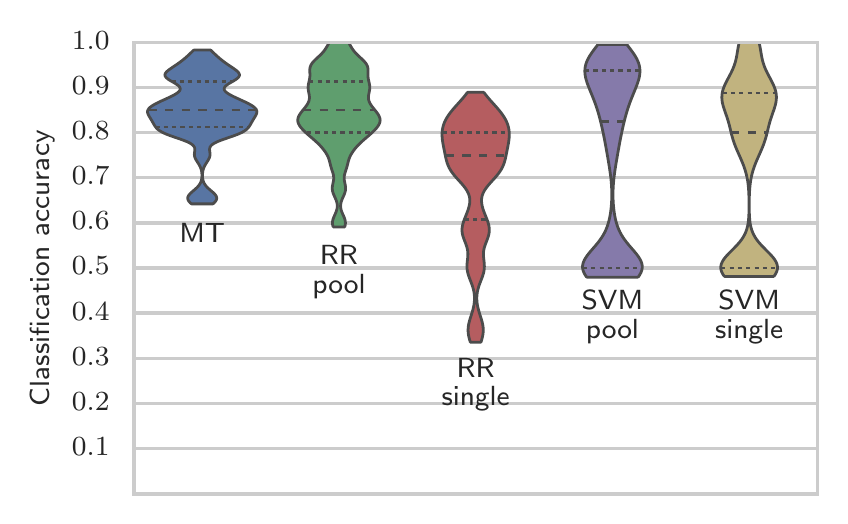}
\caption{Density plot of classification accuracy over sessions for each algorithm. MT corresponds to multitask learning using Algorithm \ref{alg:FDalgo} and RR corresponds to ridge regression using the FD regression method. Dashed line corresponds to median for the distribution and dotted lines show upper and lower quartiles. Classification accuracies using the pooled FD regression and multitask learning have a higher minimum classification accuracy than any other method. \label{Fig:GHsess} }
\end{figure}

\begin{figure}
\subfloat{\includegraphics[width=4in]{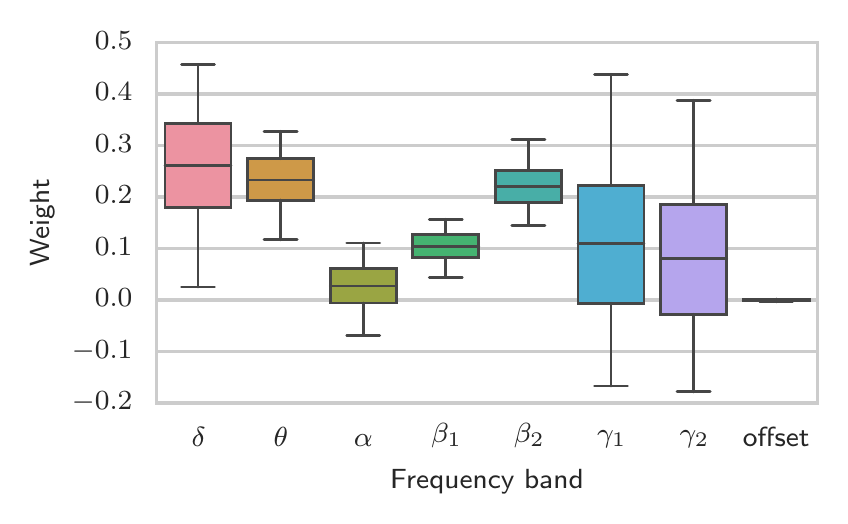}}%
\hfil
\subfloat{\includegraphics[trim=10 10 10 10,clip,width=3in]{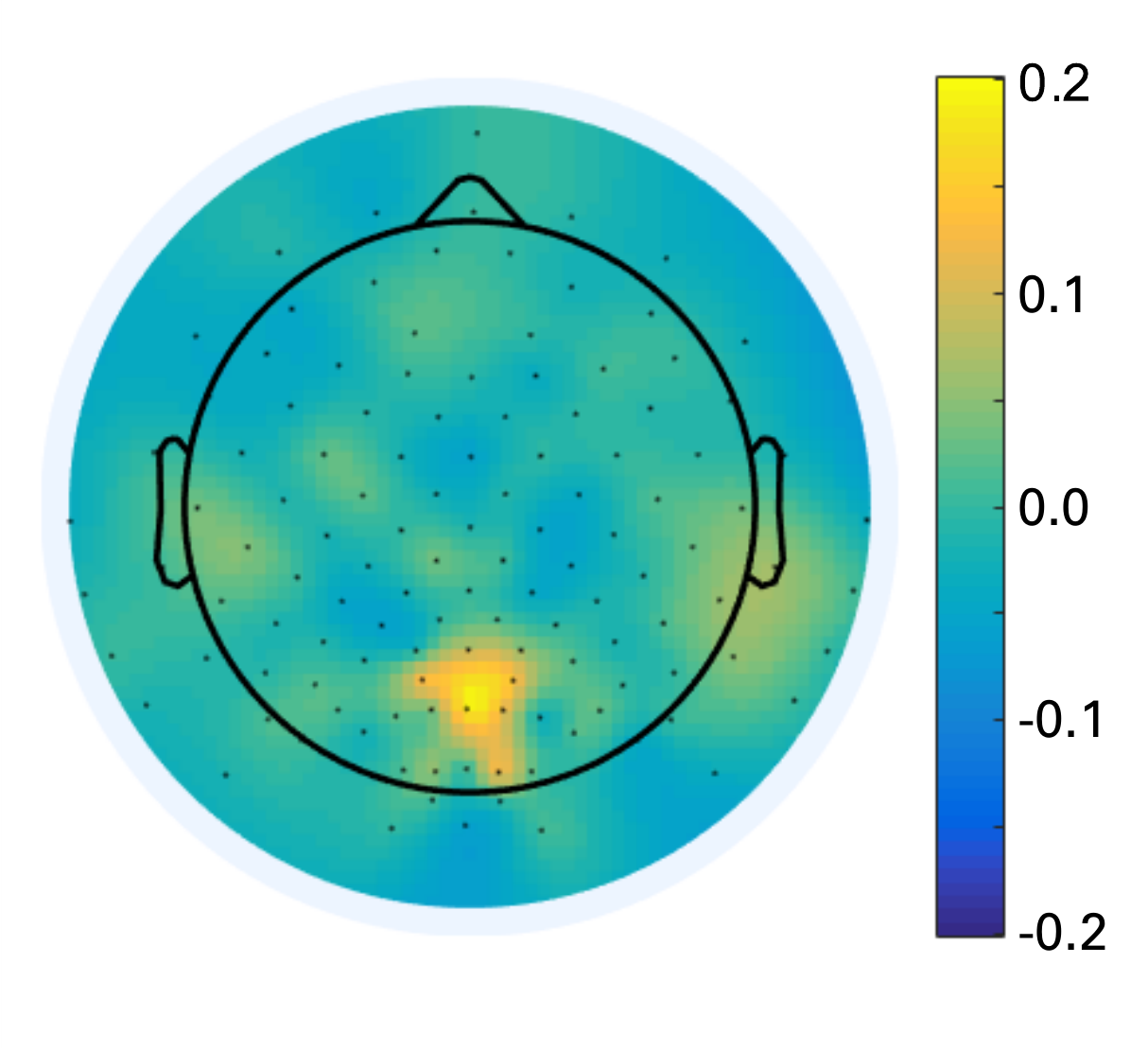}}%
\caption{\label{Fig:GHmus}(left) Box and whiskers plot with median, quartiles, and 1.5 times the IQR for frequency weights over 30 test sessions. Though this ignores the evolution of weights over time the $\delta$ range is highly weighted. (right) Sum of learned spatial mean weights
after thirty sessions plotted topographically with respect to the head, showing that the area over the parietal cortex is emphasized. }
\end{figure}

To eliminate artifacts, independent component analysis (ICA) was performed on each session using the SOBI algorithm \cite{Belouchrani1997}
and components corresponding to cortical features \cite{Onton2006,Grosse-Wentrup2013} were manually chosen. The time-series of these components
were then re-projected to the electrode space. For each trial and electrode, the log-powers in the frequency bands $\delta=1\text{-}5\ \text{Hz},\theta=5\text{-}8\ \text{Hz},\alpha=8\text{-}12\ \text{Hz},\beta_1=12\text{-}20\ \text{Hz},\beta_2=20\text{-}30\ \text{Hz},\gamma_{1}=30\text{-}70\ \text{Hz},\gamma_{2}=70\text{-}90\ \text{Hz}$ were computed
using a discrete Fourier transform over the sixty seconds of the trial
to create a $121 \times 7$ feature space. The first session was used for training, after which
the first run of each session was used to update the classifier according to Section \ref{sec:online}
and the updated weight vectors used to classify the data in the next one or two runs in a pseudo-online fashion.
Between sessions Algorithm \ref{alg:FDalgo} was re-run with all data of the most recent session included, as we found experimentally that 
the non-initialized case performed better on these data. We compare results between the MT, RR, and SVM performance (Figure \ref{Fig:GHsess}). The 
spatial and frequency weights learned by the MT algorithm are shown in Figure \ref{Fig:GHmus}. Single and pooled were computed similarly to those presented in Section \ref{subsec:classperf} except that instead of subjects we used sessions. 

\subsubsection{Results}
We can see that the multitask and
the pooled ridge regression have the highest median (85\%) and show more density in higher classification accuracies. Both are significantly better than the single-session ridge-regression ($p < 0.0001$, Wilcox signed-rank test); as the SVM results are clearly bimodal a median comparison is not informative. Between the pooled and multitask FD conditions the differences are small, which
may reflect the fact that inter-session differences are not as large as inter-subject differences. However, the multitask formulation has a higher minimum
classification accuracy (65\% vs 60\%) than the pooled accuracy, suggesting that considering the sessions separately still adds a small benefit
when attempting to test on sessions that are outliers for some reason. This may be related to why the SVM distributions are bimodal, as the SVM either classifies excellently or at chance level in both the single-session and pooled cases. This also suggests that there are outlier sessions, in which the distribution of data in the feature domain is sufficiently different from past data to cause the cross-validation over the training data to poorly predict test data classification. Possibly the fact that there is no distinction made between sessions in the pooled case causes these methods to have lower minimum accuracies. Looking further to the spatial and spectral weights, we see that the weights are concentrated
in the area directly above the precuneus. Instead of a smooth topography, however, we see that certain channels are strong and nearby channels are not,
which is consistent with the feature selection aspect of the regularization as discussed in Section \ref{sec:offline}. 

%% file: Discussion.tex
\section{Discussion}
\label{sec:discussion}

We have introduced a framework for transfer learning, both across subjects and across sessions, 
that works across feature spaces and requires fewer training trials than other state-of-the-art
methods for classification, representing an effective combination of pooled data and single-subject/session training. Previous work in transfer learning for BCI focuses on transforming the
feature space of individual subjects/sessions such that one decision
boundary generalizes well across subjects/sessions, here referred to as 'domain
adaptation'. In contrast, our method treats the decision boundary as a
random variable whose distribution is inferred from previous
subjects/sessions. As a result, our method is complementary to domain
adaptation methods. Further, we show that applying this formulation with an altered regression method that takes feature decomposition
into account is effective at learning structure between both multiple subjects and multiple sessions in EEG-based BCI tasks. By assuming that the weights of the channels and the features are independent we are able to drastically reduce the size of the feature space. This method works better than an SVM trained on an equal feature space both in the zero-training transfer learning case and after a within-session training period. The prior parameters
describing the distribution of the weight vectors can also be quickly used to see spatial and spectral topographies associated with a given task.

Though the proposed regression method appears to work well across datasets, it has some undesirable features. One such characteristic of the proposed regression method is the variable importance of initializing the spatial weights smartly.
In the motor imagery paradigm, a lack of proper initialization resulted in very poor results; conversely, in the other experiment, using initialization
was less effective. While there is no clear rule as to when it might be necessary, we can easily see a possible explanation for this problem when
considering the regression method itself:
\begin{equation}
    \hat{y}=\aalpha^{\text{T}}X\w=(C\aalpha)^{\text{T}}X\left(\frac{1}{C}\w\right)
\end{equation}
where $C$ is an arbitrary real number. This symmetry means that the likelihood function is not actually convex, making the location at which it is initialized in its domain
important to the predictivity of the results. When initialized poorly, it can fail to find
predictive features. Further work may determine an appropriate criterion to make the regression method properly convex. For practical application,
however, we found no obvious trend as to which paradigms work better with a non-informative initialization versus a ridge-regression initialization. Our suggestion with this method would be to test both empirically and choose the one that works best.

A second problematic result of using the FD regression is the addition of another loop in the algorithm, as now for each subject/session there must be
iteration to determine an appropriate spatial and spectral combination. However, in practice we found this to run quite smoothly. The other option
is to use the regular regression method, which results in a far larger matrix that has to be inverted for every session. We also found that
the convergence in the case of the FD regression happened orders of magnitude faster than in the non-FD case, possibly due to the far more favorable
ratio of training trials to features. Overall, though there is a second loop in the algorithm, the FD case is actually faster than the non-FD case, in practice, on high-dimensional datasets.
Finally, we note that the restriction of a single spatial weight vector and frequency weight vector means that a single brain process can be classified
at a time. Winning entries in the BCI competition IV mostly used multiple signals to achieve their high accuracies \cite{Tangermann2012}, a possibility
that is not possible using this approach as they would return conflicting regression weights.

Though ours is the first presentation of inference for the full distribution of weight vectors in BCI, this approach has been well-studied
in the machine learning domain for a variety of different problems \cite{Yu2005}. One possible future direction is to specify our priors as samples from a Dirichlet process and attempt to take advantage of any clustering as the number of subjects increases \cite{Xue2007}, as has been shown to be effective in CSP multi-subject learning \cite{Kang2012}. It is also interesting to note that the multitask learning formulation is simply an additive convex term to the loss function, which suggests that it can be added to any algorithm as a cheap way of learning something about the shared structure of classification rules (though without some involvement of the shared parameters in the computation of the task-specific rules an iterative procedure would be impossible). Further work with this approach in SVMs or LDA should prove to be very interesting.
Lastly, the current approach requires that the entire iterative scheme is re-run after the inclusion of any new subjects or sessions, which quickly becomes
inefficient as the number of considered subjects or sessions increases. More research to help streamline the update rule of the priors would be invaluable
in the age of big data. 

It is likely that all the methods presented here would perform better if prior knowledge had been incorporated into choosing the feature space. For example,
Alamgir et al.~\cite{Alamgir2010} use data only from the electrodes directly over the motor cortex. Indeed, given a small feature space and a separable
problem, it is well known that optimizing the objective function of an SVM leads to better test classification than simple least-squares loss. The problem is simply
that we do not always have so much prior information; further, in the case of newer paradigms such as the one the ALS patient was trained on, such information
is currently unavailable, a problem that will only continue as more possible paradigms are experimented with. We hope that the multitask framework presented here will function as a way of quickly judging the efficacy and activation topography of new BCI paradigms. By training with feature decomposition we are able to get a picture of what channels and features are important to the task
at hand, and can then possibly re-run with the non-FD algorithm in order to better capture the multitask structure in the smaller feature space. However, there are also instances in which the data has a very large number of dimensions that do actually contribute to the classification of the task at hand, and
we have shown that multitask learning is robust to these sorts of datasets.